\documentclass[aps,amsmath,showpacs,showkeys,superscriptaddress,10pt]{revtex4}
\usepackage{color}
\usepackage{amssymb}
\usepackage{titlesec}
\usepackage{amssymb,amsfonts,amsthm}
\begin{document}
\title{Inverse variational problem for nonlinear dynamical systems}
\author{Basir Ahamed Khan}
\affiliation{Department of Physics, Krishnath College, Berhampore, Murshidabad 742101, India}
\author{Supriya Chatterjee}
\email{supriya_2k1@rediffmail.com}
\affiliation{Department of Physics, Bidhannagar College, EB-2, Sector-1, Salt Lake, Kolkata 700064, India}
\author{Sekh Golam Ali} 
\affiliation{Department of Physics, Kazi Nazrul University, Asansol 713303, India}
\author{Benoy Talukdar}
\affiliation{Department of Physics, Visva-Bharati University, Santiniketan 731235, India}
\begin{abstract}
In this paper we have chosen to work with two different approaches to solving the inverse problem of the calculus of variation. The first approach is based on an integral representation of the Lagrangian function that uses the first integral of the equation of motion while the second one relies on a generalization of the well known Noether's theorem and constructs the Lagrangian directly from the equation of motion. As an application of the integral representation of the Lagrangian function we first provide some useful remarks for the Lagrangian of the modified Emden-type equation and then obtain results for Lagrangian functions of (i) cubic-quintic Duffing oscillator, (ii) Li\'{e}nard-type oscillator and (iii) Mathews-Lakshmanan oscillator. As with the modified Emden-type equation these oscillators were found to be characterized by nonstandard Lagrangians except that one could also assign a standard Lagrangian to the Duffing oscillator. We used the second approach to find indirect analytic (Lagrangian) representation for three velocity-dependent equations for (iv) Abraham-Lorentz oscillator, (v) Lorentz oscillator and (vi) Van der Pol oscillator. For each of the dynamical systems from (i)-(vi) we calculated the result for Jacobi integral and thereby provided a method to obtain the Hamiltonian function without taking recourse to the use of the so-called Legendre transformation.
\end{abstract}
\pacs{02.30.Hq, 02.30.Ik, 02.40.Yy, 45.20.J}
\keywords{Lagrangians, Jacobi integrals, Hamiltonians, Nonlinear Differential Equations}
\maketitle
\section*{1. Introduction}
The inverse problem in the calculus of variation involves deciding whether a given system of second-order ordinary differential equations representing dynamical systems is a solution of the Euler-Lagrange equation and eventually, finding its Lagrangian representation if the solution exists \cite{1}.  For linear ordinary differential equations, the set of necessary and sufficient conditions for the existence of Lagrangians is provided by the so-called Helmholtz conditions \cite{2}. The equation of motion of a damped Harmonic oscillator
\begin{equation}
\ddot{x}(t)+\gamma\dot{x}(t)+\omega^2x(t)=0
\end{equation}
violates these conditions such that we cannot find a time-independent Lagrangian representation for it. In Eq.(1) over-dots denote differentiation with respect to $t$. We shall follow this convention throughout. Here $\gamma$ represents the frictional coefficient of the medium in which the oscillator of angular frequency $\omega$ is embedded. An explicitly time-dependent Lagrangian of the damped system was, however, found \cite{3,4} during 1940's. For this Lagrangian the canonical momentum is time dependent. This provides an awkward analytical constraint to use the corresponding Hamiltonian to quantize the system \cite{5}. In this context we note that, as early as 1931, Bateman \cite{6} suggested a very ingenuous method to find an explicitly time-independent Lagrangian for the damped Harmonic oscillator by doubling the number of degrees of freedom of the system. More specifically, in conjunction with Eq. (1), an auxiliary oscillator equation
\begin{equation}
\ddot{y}(t)-\gamma\dot{y}(t)+\omega^2y(t)=0
\end{equation}
was considered to obtain the Lagrangian
\begin{equation}
L=\dot{x}(t)\dot{y}(t)+\frac{\gamma}{2}(x(t)\dot{y}(t)-\dot{x}(t)y(t))-\omega^2x(t)y(t).
\end{equation}
Physically, the energy drained out from the oscillator in Eq. (1) is completely absorbed by that in Eq. (2) such that these two oscillators taken together represent a conservative system. It is of interest to note that Euler-Lagrange equation \cite{7} written in terms of $y(t)\;(x(t))$ gives the equation of motion for $x(t)\;(y(t))$. Because of this unusual behavior, the Lagrangian in Eq. (3) is said to provide an indirect analytic (Lagrangian) representation of the system. However, canonical quantization of damped Harmonic oscillator using the indirect Lagrangian representation has been found to be quite satisfactory \cite{8,9,10} because the corresponding Hamiltonian is time independent.
\par Traditionally, the Lagrangian function $L$ of an autonomous differential equation is expressed as $L=T-V$ where $T$ and $V$ stand for the kinetic and potential energies of the system represented by the equation. Such a Lagrangian is referred to as standard. Relatively recently, a new type of Lagrangian functions \cite{11} have been proposed for dissipative-like autonomous differential equations. These do involve neither $T$ nor $V$. As a result such Lagrangians were qualified as nonstandard. One can also find nonstandard Lagrangian representation for linear differential equations which violate of Helmholtz conditions \cite{12}. However, during the end of last decade Musielak \cite{13}, and Cie\'{s}li\'{n}ski and Nikiciuk \cite{14} presented methods to write results for nonstandard Lagrangians of a variety of nonlinear dynamical systems. It was observed that some of the equations follow for inverse type of Lagrangian functions while others follow logarithmic functions. For example, the modified Emden-type equation \cite{15}
\begin{equation}
\ddot{x}+\alpha x\dot{x}+\beta x^3=0,\;\;\;\;x=x(t)
\end{equation}
was found to have Lagrangian representations given by \cite{11}
\begin{subequations}
\begin{equation}
 L=\frac{1}{\dot{x}+kx^2}\;\;\;\;\mbox{for}\;\;\;\; \alpha=3k\;\;\;\mbox{and}\;\;\;\; \beta=\alpha^2/9
\end{equation}
 \mbox{and}
\begin{equation}
L=\ln(\dot{x}+kx^2)\;\;\;\;\mbox{for}\;\;\;\; \alpha=4k\;\;\;\mbox{and}\;\;\;\; \beta=\alpha^2/8.
\end{equation}
\end{subequations}
The results for Lagrangians of Eq. (4) and of similar equations were chosen in an ad hoc fashion rather than being derived from the solution of the inverse problem in the calculus of variation \cite{16}. In this context we note that the work of Nucci and Tamizhmani \cite{17} is a step forward along this line of investigation because they provide an elegant method to find Lagrangians of second-order differential equations (linear or nonlinear) by the use of Jacobi's last multiplier \cite{18}. The object of the present paper is to introduce two uncomplicated methods to solve the inverse variational problem for nonlinear differential equations and then make use of them to construct nonstandard Lagrangian representations for a number of evolution equations that play a role in many applicative contexts. The first method of our interest makes use of an integral representation of the Lagrangian function in terms of the first integral of the equation of motion \cite{19}. The second method sought by Hojman \cite{20} depends on a simple but nontrivial generalization of the famous theorem of Noether \cite{21}.
\par The Lagrangian mechanics begins with the pre-supposition that the mechanical state of a system can be described by specifying its generalized coordinates and velocities. On the other hand, the so-called Hamiltonian formulation \cite{7} of classical mechanics provides a description of physical systems using generalized coordinates and momenta. In the case of standard Lagrangians, the traditional recipe for transition from Lagrangian to Hamiltonian formulation of mechanics consists in defining a canonical momentum and then takes recourse to the use of the Legendre transformation. We shall show that the same prescription also holds for writing Hamiltonians from nonstandard Lagrangians.
\par  In section 2 we provide a brief derivation of the methods of our interest to solve the inverse problem in point mechanics. As a useful application of the first method we demonstrate that the Lagrangian for the equation in (4) can be found for arbitrary values of $\alpha$ and  $\beta$ such that the results in Eq. (5a) and Eq. (5b) are only special instances. Further, we show that for $\alpha=3k$ and $\beta=\alpha^2/9$ one can have a logarithmic Lagrangian representation of Eq. (4) in addition to the inverse one as given in Eq. (5a). But Eq. (4) for $\alpha=4k$ and $\beta=\alpha^2/8$ does not have any inverse type Lagrangian. As regards the application of the second method we provide an ab intio derivation of the Bateman Lagrangian in Eq. (3) and find that the approach is especially important for velocity-dependent differential equations. We devote section 3 to present an uncomplicated method to find the first integral or constant of the motion of nonlinear differential equations and thereby obtain Lagrangian representations of number dynamical systems using our first method. Where-ever possible we present results for both inverse and logarithmic type of Lagrangians. Here we also implement the second method of our interest to find Lagrangians for two nonlinear dissipative equations and one third-order equation. In section 4 we look for Jacobi integrals \cite{7} for the nonstandard Lagrangians found in section 3 and then provide results for the corresponding Hamiltonian functions. And finally, in section 5 we summarize our outlook on the present work and make some concluding remarks. 
\section*{2. Methods for constructing of Lagrangian representation}
Here we present methods of our interest to solve the inverse problem of the calculus of variations. These approaches are particularly well-suited for constructing non-standard Lagrangian representation of nonlinear evolution equations.
\par {\bf (i) Method 1: Relation between Lagrangian and constant of the motion}
\par The relationship between the Lagrangian and constant of the motion is represented by the Jacobi integral \cite{7}
\begin{equation}
\sum_{i=1}^Nv^i(\frac{\partial L}{\partial v^i})-L=K
\end{equation}
where $L=L(\vec{x},\vec{v})$ is the Lagrangian and $K=K(\vec{x},\vec{v})$, the constant of the motion of the second-order ordinary differential equation
\begin{equation}
\ddot{x}^i=f^i(x^j,\dot{x}^j),\;\;\;\;i=1,\;2,........,N.
\end{equation}
Here $\vec{x}=(x^1,.....,x^N)$  and $\vec{v}(=\vec{\dot{x}})=(v^1,.....,v^N)$. The equations for the characteristics of Eq. (6) are \cite{22}
\begin{equation}
\frac{dv^1}{v^1}=.....=\frac{dv^N}{v^N}=\frac{dL}{L+K}=\frac{dx^1}{0}=......=\frac{dx^N}{0}.
\end{equation}
Writing Eq. (7) in the equivalent form $\frac{dv^i}{dt}=f^i(\vec{x},\vec{v})$, one can demand that the constant of the motion $K(\vec{x},\vec{v})$ is a first integral of the equation provided
\begin{equation}
\sum_{i=1}^N[f^i(\vec{x},\vec{v})\frac{\partial K}{\partial v^i}+v^i\frac{\partial K}{\partial x^i}]=0.
\end{equation}
The solution of Eq. (9) or the integral surfaces can now be obtained from the equation of characteristic 
\begin{equation}
\frac{dv^1}{f^1(\vec{x},\vec{v})}=......=\frac{dv^N}{f^N(\vec{x},\vec{v})}=\frac{dx^1}{v^1}=.....=\frac{dx^N}{v^N}=\frac{dK}{0}.
\end{equation}
The last term in Eq. (10) clearly indicates that it can be used to find the general solution of Eq. (9), which represents a constant of the second-order differential equation (7). For an $N$-dimensional autonomous Newtonian system the solution of Eq. (7) can be expressed as an integral over possible constants of the motion \cite{19} and we have
\begin{subequations}
\begin{equation}
L(\vec{x},\vec{v})=\frac{1}{N}\sum_{i=1}^Nv^i\int^{v^i}\frac{K^i(\vec{x},\xi)}{\xi^2}d\xi.
\end{equation}
\mbox{For the one-dimensional system Eq. (11a) reads}
\begin{equation}
L(x,v)=v\int^{v}\frac{K(x,\xi)}{\xi^2}d\xi.
\end{equation}
\end{subequations}
In the following we shall make use of the integral representation of the Lagrangian function in Eq. (11) to deal with the inverse problem of Eq. (4) in a more general context. To find the constant of the motion we shall follow a very simple prescription rather than taking recourse to the use of Eqs. (9) and (10).
\par We begin by writing Eq. (4) in the autonomous form
\begin{equation}
v(x)v'(x)+\alpha xv(x)+\beta x^3=0,
\end{equation}
where $v(x)=\frac{dx}{dt}$ and $v'(x)=\frac{dv(x)}{dx}$. This first-order linear differential equation can be solved to get a constant of the motion
\begin{equation}
K(x,\dot{x})=\frac{2\alpha\arctan h[\frac{\alpha x^2+4\dot{x}}{x^2\sqrt{\alpha^2-8\beta}}]}{\sqrt{\alpha^2-8\beta}}+\ln[\beta x^4+\alpha x^2\dot{x}+2x^2]
\end{equation}
In principle, one can use Eq. (13) in Eq. (11b) to construct a Lagrangian for the modified Emden-type equation. But it will be instructive to write, from Eq. (13), a constant of the motion for $\alpha=3k$ and $\beta=\alpha^2/9$, and find for Eq. (5a) two Lagrangian representations, which are not related by a trivial gauge term. For these special values of the parameters, the constant of the motion reduces to a very simple form given by
\begin{equation}
K(x,\dot{x})=\frac{(kx^2+\dot{x})^2}{kx^2+2\dot{x}}.
\end{equation}
If $K(x,\dot{x})$ is a constant of the motion, its reciprocal 
\begin{equation}
K_1(x,\dot{x})=1/K(x,\dot{x})
\end{equation}
is also a constant of the motion. It is straightforward to verify that $K_1(.)$ in conjunction with Eq. (11b) leads to the reciprocal Lagrangian as appears in Eq. (5a). Similarly, using Eq. (14) we obtain a logarithmic-type Lagrangian
\begin{equation}
L_1=\frac{1}{2}\dot{x}\ln(kx^2+2\dot{x})-kx^2.
\end{equation}
The Lagrangians in Eqs. (5a) and (16) for the same equation of motion are not related by a gauge term. Such Lagrangians are called alternative or in-equivalent Lagrangians \cite{22}. The existence of alternative Lagrangian description of physical systems has important consequences for the correspondence between symmetries and constants of the motion \cite{23}.
\vskip 0.25cm
\par {\bf (ii) Method 2: Construction of Lagrangians using the equations of motion}
\par While recognizing the importance of first integral in solving inverse variational problem, Hojman et al \cite{24} raised a very important question. Can the equations of motion themselves, rather than their first integrals, be used to provide Lagrangian description of mechanical systems? In \cite{20} it was firmly established that for linear Newtonian systems one can always use the equations of motion to find a satisfactory solution of the inverse problem. We shall show that the method can also be adapted to deal with nonlinear problems. The basic philosophy of the method can be understood as follows.
\par  It is well known in classical mechanics that the so-called Noether's theorem \cite{21} provides a relation between symmetries of the Lagrangian with conserved quantities of the equation of motion. But it is less well known that the symmetries of the equations of motion form a larger set than the symmetries of the Lagrangians. However, if s-equivalence is taken into account, the set of Lagrangian symmetries coincides with that of the equation of motion. By s-equivalence we mean a Lagrangian symmetry in which several constants of the motion may be associated with one symmetry transformation \cite{25}. Understandably, the work in \cite{24} is a generalization of the traditional Noetherian symmetry. In this context an interesting result that was found is outlined below.
\par If the second-order differential equation (7) follows from an action principle with the Lagrangian $L(x^i,\dot{x}^i)$, then this equation will also follow from a higher-order action characterized by an acceleration-dependent Lagrangian $\bar{L}(x^i,\dot{x}^i,\ddot{x}^i)$ given by 
\begin{equation}
\bar{L}(x^i,\dot{x}^i,\ddot{x}^i)=\mu_i(\ddot{x}^i-f^i(x^i,\dot{x}^i)).
\end{equation}
The Lagrangians $L(.)$ and $\bar{L}(.)$ are related by a gauge term written as
\begin{equation}
L(x^i,\dot{x}^i)=\bar{L}(x^i,\dot{x}^i,\ddot{x}^i)+\frac{dg(x^i,\dot{x}^i)}{dt}
\end{equation}
such that
\begin{equation}
\mu_i=-\frac{\partial g(x^i,\dot{x}^i)}{\partial\dot{x}^i}.
\end{equation}
In general, the concept of higher-order action that leads to a generalized classical mechanics is due to Euler who found the differential equation \cite{26,27}
\begin{equation}
\sum_{i=0}^n(-1)^i(\frac{d}{dt})^i\frac{\partial L}{\partial x^{(i)}}=0
\end{equation}
for the n-th order Lagrangian $L=L(x^i,\dot{x}^i,\ddot{x}^i,......,x^{(n)})$. We shall now show that the auxiliary equation (2) introduced by Bateman \cite{6} to find a time-independent Lagrangian for the damped Harmonic oscillator follows naturally from Eq. (17) and thereby obtain the result in Eq. (3).
\par From Eq. (1) and Eq. (17) we write
\begin{equation}
\bar{L}(.)=y(\ddot{x}+\gamma\dot{x}+\omega^2x)
\end{equation}
and substitute Eq. (21) in the second-order Euler-Lagrange equation obtained from Eq. (20) for $n=2$. This gives the expected equation
\begin{equation}
\ddot{y}-\gamma\dot{y}+\omega^2y=0.
\end{equation}
In view of (18) we write the second-order Lagrangian for the uncoupled systems in Eqs. (1) and (2) as
\begin{equation}
L(.)=y(\ddot{x}+\gamma\dot{x}+\omega^2x)+x(\ddot{y}-\gamma\dot{y}+\omega^2y)-\frac{d}{dt}(y\dot{x}+x\dot{y}).
\end{equation}
The third term in Eq. (23) stands for the gauge term of the second-order Lagrangian \cite{28}.
\section*{3. Lagrangian description of nonlinear dynamical systems}
Here we shall make use of methods 1 and 2 to compute results for some typical nonlinear systems. In particular, method 1 will be employed to find Lagrangian representation of (i) cubic-quintic Duffing oscillator \cite{29}, (ii) Li\'{e}nard-type nonlinear oscillator or generalized Emden-type equation \cite{15} and (iii) Mathews-Lakshmanan oscillator \cite{30}. On the other hand, similar results for (iv) Abraham-Lorentz  oscillator \cite{31}, (v) Lorenz oscillator \cite{32} and (vi) Van der Pol oscillator \cite{33} will be obtained by the use of method 2. Here we must point out that the oscillator in (iv) is represented by a third-order linear differential equation. Our reason for providing an ab initio derivation for the Lagrangian function $L_{al}(.)$ for (iv) is that $L_{al}(.)$ plays a crucial role in quantizing the radiation-damped Harmonic oscillator \cite{34}.
\vskip0.25cm
\par (i) The cubic-quintic Duffing oscillator represented by 
\begin{equation}
 \ddot{x}+ax+bx^3+cx^5=0
\end{equation}
with constant values of $a$, $b$ and $c$ arise in a number of applicative contexts \cite{35}. The first-order autonomous differential equation corresponding to Eq. (24) given by
\begin{equation}
 v(x)v'(x)+ax+bx^3+cx^5=0
\end{equation}
can easily be integrated to get the constant of the motion
\begin{equation}
 K(x,\dot{x})=6ax^2+3bx^4+2cx^6+6\dot{x}^2.
\end{equation}
From Eqs. (11b) and (26) we find the Lagrangian 
\begin{equation}
 L=\frac{1}{2}\dot{x}^2-\frac{1}{2}ax^2-\frac{1}{4}bx^4-\frac{1}{6}cx^6.
\end{equation}
A number of inequivalent Lagrangians can be constructed by using various powers of the first integral in Eq. (27). However, in close analogy with our observation on the results of the modified Emden-type equation (4), it may be of some interest to find the Lagrangian of Eq. (24) for $K_1(x,\dot{x})$. In this case we get a fairly complicated result 
\begin{equation}
 L_1=\frac{1}{rx^2}+\frac{\sqrt{6}\dot{x}}{x^3r^{3/2}}\arctan(\frac{\sqrt{6}\dot{x}}{\sqrt{r}x})
\end{equation}
with
\begin{equation}
 r=6a+3bx^2+2cx^4.
\end{equation}
The cubic-quintic Duffing oscillator is primarily represented by the standard Lagrangian (27). But Eq. (28) shows that it can also be represented by a nonstandard Lagrangian.
\vskip0.25cm
\par (ii) Here the equation of our interest is the Li\'{e}nard-type nonlinear oscillator represented by \cite{15}
\begin{equation}
 \ddot{x}+kx\dot{x}+\lambda x+\frac{k^2x^3}{9}=0.
\end{equation}
While studying the dynamical properties of Eq. (30), Chandrasekar et al \cite{36} provided a non-standard Lagrangian representation for the equation. We shall show that in respect of this a relatively simpler representation can be obtained by using the integral representation (11b). For Eq. (30), a first-order differential equation analogous to that in Eq. (25) reads
\begin{equation}
 v(x)v'(x)+kxv(x)+\lambda x+\frac{k^2x^3}{9}=0.
\end{equation}
The first integral of the Li\'{e}nard-type equation obtained from Eq. (31) is given by
\begin{equation}
 K(x,\dot{x})=\frac{(9\lambda+3k\dot{x}+k^2x^2)^2}{9\lambda+6k\dot{x}+k^2x^2}.
\end{equation}
From Eqs. (11b) and (32) we now obtain the Lagrangian
\begin{equation}
 L=3k\dot{x}\ln(9\lambda+6k\dot{x}+k^2x^2)-2k^2x^2.
\end{equation}
Similarly, for $K_1(x,\dot{x})=1/K(x,\dot{x})$ we found
\begin{equation}
 L_1=\frac{1}{9\lambda+3k\dot{x}+k^2x^2}.
\end{equation}
Interestingly, from Eqs. (33) and (34) we see that as with Eq. (4) ($\alpha=3k$, $\beta=\alpha^2/9$), the Li\'{e}nard-type equation also can have both logarithmic and inverse Lagrangian representation. 
\vskip0.25cm
\par (iii) The Mathews-Lakshmanan oscillator \cite{30}
\begin{equation}
 \ddot{x}+\frac{\lambda x}{1-\lambda x^2}\dot{x}^2+\frac{\omega^2x}{1-\lambda x^2}=0
\end{equation}
may be regarded as the zero-dimensional version of a scalar non-polynomial field equation \cite{37}. It can also be considered as an oscillator with position-dependent effective mass \cite{38}. Writing Eq. (35) as a first-order autonomous differential equation we have found its first integral as
\begin{equation}
 K(x,\dot{x})=\frac{\lambda\dot{x}^2-\omega^2}{1+\lambda x^2}.
\end{equation}
Equation (36) in conjunction with Eq. (11b) leads to well known result for the Lagrangian \cite{39}
\begin{equation}
 L=\frac{\lambda\dot{x}^2+\omega^2}{1+\lambda x^2}.
\end{equation}
Similarly, the constant of the motion $K_1(x,\dot{x})=1/K(x,\dot{x})$ gives
\begin{equation}
 L_1=(1+\lambda x^2)(\omega-\sqrt{\lambda}\dot{x}\arctan h(\frac{\sqrt{\lambda}\dot{x}}{\omega})).
\end{equation}
Since $\arctan hx=\frac{1}{2}\ln(\frac{1+x}{1-x})$, the fairly complicated result in Eq. (38), in fact, represents a logarithmic Lagrangian.
\vskip0.25cm
\par (iv) The motion of an accelerated charged particle including the reactive effects is given by the so-called  Abraham-Lorentz equation \cite{31} which in one dimension can be written as
\begin{equation}
 m\ddot{x}+kx-m\tau\dddot{x}=0.
\end{equation}
Here $m$ stands for the mass of the electron and $k$, the spring constant of its Harmonic motion. The third term in Eq. (39) 
 with
\begin{equation}
 \tau=\frac{2}{3}\frac{e^2}{mc^2}
\end{equation}
has its origin in the radiative reaction and represents the damping term in the equation of motion. We shall present here an indirect Lagrangian representation of Eq. (39). To that end we write a Lagrangian 
\begin{equation}
 \bar{L}= y(m\ddot{x}+kx-m\tau\dddot{x})
\end{equation}
characterized by two degrees of freedom, namely, $x(t)$ and $y(t)$. The third-order Lagrangian when substituted in Eq. (20) for $n=3$ gives the associated equation \cite{6}
\begin{equation}
m\ddot{y}+ky+m\tau\dddot{y}=0 
\end{equation}
such that we can write an indirect Lagrangian 
\begin{equation}
 L=y(m\ddot{x}+kx-m\tau\dddot{x})+x(m\ddot{y}+ky+m\tau\dddot{y})+a-b
\end{equation}
for the dual system. Here $a$ and $b$ given by
\begin{equation}
 a=m\tau\frac{d}{dt}(y\ddot{x}-x\ddot{y})
\end{equation}
and 
\begin{equation}
 b=m\frac{d}{dt}(y\dot{x}+x\dot{y})
\end{equation}
represent the appropriate gauge terms which have been used to write Eq.(43) in the maximally reduced form
\begin{equation}
 L=m\dot{x}\dot{y}+\frac{1}{2}m\tau(\dot{x}\ddot{y}-\dot{y}\ddot{x})-kxy.
\end{equation}
From Eq. (46) we see that we need a second-order Lagrangian to analytically represent the third-order equation (39). As expected in the absence of radiative reaction Eq. (46) i.e. $\tau=0$ gives the usual Harmonic oscillator Lagrangian. 
\vskip0.25cm
\par (v)  The simple nonlinear differential equation \cite{40}
\begin{equation}
 \ddot{x}+\delta\dot{x}-\mu x+\alpha x^2=0
\end{equation}
with constant values of $\alpha$, $\delta$ and $\mu$ represent the Helmholtz oscillator. Since it involves a dissipative term linear in velocity, its  Lagrangian representation can be found by following the method used in the previous example. But here we should remember that the linear and nonlinear terms contribute to the Lagrangian with unequal weights \cite{41}. The weight factors are determined by demanding that the computed Lagrangian function should reproduce the equation of motion via the Euler-Lagrange equation. In analogy with Eq. (41) we thus write
\begin{equation}
 \bar{L}=y(\ddot{x}+\delta\dot{x}-\mu x+w\alpha x^2),
\end{equation}
where $w$ stands for the required weight factor. In writing Eq. (48) we assumed that weight factor of the linear terms is unity. The postulated  second-order Lagrangian in Eq. (48) leads to the associated equation
\begin{equation}
 \ddot{y}-\delta\dot{y}-\mu y+2w\alpha xy=0
\end{equation}
that we need to construct $L$. We thus write the Lagrangian for Eq. (47) in the form
\begin{equation}
 L=y(\ddot{x}+\delta\dot{x}-\mu x+w\alpha x^2)+x(\ddot{y}-\delta\dot{y}-\mu y+2w\alpha xy)-\frac{d}{dt}(y\dot{x}+\dot{y}x).
\end{equation}
We have verified that $L$ in Eq. (50) when used in the Euler-Lagrange equation reproduces Eq. (47) if $w=2/3$. Our final result for the required Lagrangian reads
\begin{equation}
 L=\dot{x}\dot{y}+\frac{1}{2}\delta(x\dot{y}-y\dot{x})-\alpha x^2y+\mu xy.
\end{equation}
\vskip0.25cm
\par (vi) The Van der Pol equation
\begin{equation}
 \ddot{x}-\mu(1-x^2)\dot{x}+x=0
\end{equation}
for a positive parameter $\mu$ represents an oscillator which absorbs energy from the surrounding when $x<1$ and dissipates energy when $x>1$. Periodic motion of this type is qualified as relaxation oscillation. A great variety of physical processes ranging from economic crisis to beating of the human heart can be modeled by Eq. (1). Following the method used for the Helmholtz oscillator we have found the Lagrangian 
\begin{equation}
 L=\dot{x}\dot{y}+\frac{1}{2}\mu(\dot{x}y-\dot{y}x)+\frac{1}{4}\mu(x^3\dot{y}-x^2y\dot{x})-xy
\end{equation}
which can reproduce both Eq. (52) and its associate equation given by
\begin{equation}
 \ddot{y}+\mu(1-x^2)\dot{y}+y=0.
\end{equation}
\section*{4. Hamiltonizing nonstandard Lagrangians}
We shall present here a form of analytical dynamics for nonlinear equations, which is not particularly superior to Lagrangian mechanics for treating classical problems but it provides a framework for theoretical extension to other advanced areas of physics including quantum mechanics \cite{42}. This alternative statement for the structure of mechanics goes by the name Hamiltonian formulation of classical mechanics. For historical reason we shall begin with the so-called Jacobi integral which provides a statement for the conservation of energy \cite{18}. For a first-order Lagrangian $L(\dot{q}_i,q_i,t)$ involving only the generalized coordinates and velocities, the Jacobi integral 
\begin{equation}
 J^{(1)}=\dot{q}_i\frac{\partial L}{\partial\dot{q}_i}-L
\end{equation}
can be found in any standard text book. But it is not easy to find a similar result for higher-order Lagrangians characterized by higher derivative of $q_i$. However, for a second-order Lagrangian $L(\ddot{q}_i,\dot{q}_i,q_i,t)$ we have found
\begin{equation}
 J^{(2)}=\ddot{q}_i\frac{\partial L}{\partial\ddot{q}_i}+\dot{q}_i\frac{\partial L}{\partial\dot{q}_i}-\dot{q}_i\frac{d}{dt}\frac{\partial L}{\partial\ddot{q}_i}-L.
\end{equation}
It is straightforward to verify that, for the cubic-quintic Duffing oscillator, the Jacobi integrals, $J^{(1)}$'s, computed by using the Lagrangians in Eqs. (27) and (28) are in exact agreement with the constant of the motion $K(x,\dot{x})$ (Eq. (26)) and its reciprocal. Similar conclusions also hold good for the Lagrangians of the Li\'{e}nard-type oscillator and Mathews-Lakshmanan oscillator.
\par We calculated the Lagrangian functions for dissipative dynamical systems like the damped harmonic oscillator, Abraham-Lorentz, Lorentz and Van der Pol oscillator directly from their equations of motion. It will, therefore be interesting to use the Jacobi integral to find the constant of the motion for each of them and thus attain some added realism for the problem. We first calculate the Jacobi integral for the damped Harmonic oscillator by using Eq. (55) and the Lagrangian function in Eq. (23), and find
\begin{equation}
 J^{(1)}_{dho}=\dot{x}\dot{y}+\omega^2xy.
\end{equation}
The result in Eq. (57) does not have any effect of dissipation and is exactly the same as that found for two uncoupled Harmonic oscillators using their indirect analytic or Lagrangian representation \cite{43}. The reason for this is that the damped Harmonic and its associate form a conservative system. However, the time derivative of Eq. (57) can be written in the form
\begin{equation}
 \frac{dJ^{(1)}_{dho}}{dt}=\dot{y}(\ddot{x}+\gamma\dot{x}+\omega^2x)+\dot{x}(\ddot{y}-\gamma\dot{y}+\omega^2y).
\end{equation}
Since $\frac{dJ^{(1)}_{dho}}{dt}$ can be made to vanish by using the equations of motion for the damped Harmonic oscillator and its associate, the expression in Eq. (58) really represents a constant of the motion for Eq. (1). The Jacobi integrals for the Lorentz and Van der Pol oscillators are given by 
\begin{equation}
 J^{(1)}_L=\dot{x}\dot{y}+\alpha x^2y-\mu xy
\end{equation}
and
\begin{equation}
 J^{(1)}_{VP}=J^{(1)}_{dho}|_{\omega=1}
\end{equation}
respectively.
\par The Abraham-Lorentz system is described by a second-order Lagrangian given in Eq. (46). Therefore its Jacobi integral will be computed by using Eq. (56). From Eq. (46) we have 
\begin{subequations}
\begin{equation}
\frac{\partial L}{\partial\ddot{x}}=-\frac{1}{2}m\tau\dot{y},\;\;\;\frac{\partial L}{\partial\ddot{y}}=\frac{1}{2}m\tau\dot{x}
\end{equation}
\begin{equation}
\frac{\partial L}{\partial\dot{x}}=m\dot{y}+\frac{1}{2}m\tau\ddot{y}\;\;\;\mbox{and}\;\;\;\frac{\partial L}{\partial\dot{y}}=m\dot{x}-\frac{1}{2}m\tau\ddot{x}. 
\end{equation}
\end{subequations}
From Eqs. (46) and (56) we get the Jacobi integral for the Abraham-Lorentz equation as
\begin{equation}
J^{(2)}_{AL}=m\dot{x}\dot{y}+m\tau(\dot{x}\ddot{y}-\dot{y}\ddot{x})+kxy.
\end{equation}
 It is straightforward to see for the first-order Lagrangian, the Jacobi integral in Eq.(55) provides a useful basis for  smooth transition from the description of mechanical systems in $(q_i,\dot{q}_i)$ space to that in $(q_i,p_i)$ space. This can be achieved by first introducing the definition of canonical momentum
\begin{equation}
 p_i=\frac{\partial L}{\partial\dot{q}_i}
\end{equation}
and then replacing the Jacobi integral by Hamiltonian function $H$ to write
\begin{equation}
 H(q,p,t)=\{p_i\dot{q}_i-L(q,p,t)\}|_{\dot{q}=\dot{q}(q,\dot{q},t)}.
\end{equation}
It is well known that Euler-Lagrange equations result from Hamilton's principle considered in the $(q,\dot{q},t)$ space. Similarly, a variational principle in the phase space \cite{44} leads to the Hamilton's equations of motion
\begin{equation}
 \dot{q}_i=\frac{\partial H}{\partial p_i}\;\;\;\; \mbox{and}\;\;\;\;\dot{p}_i=-\frac{\partial H}{\partial q_i}.
\end{equation}
The recipe given above for going from Lagrangian to Hamiltonian holds good for the standard Lagrangian. It is of interest to verify that if Eqs. (63) and (64) are also true for nonstandard Lagrangians. To that end we make use of Eqs.(63) and (64) to obtain the Hamiltonian function 
\begin{equation}
 H=\frac{p^2}{24}+6ax^2+3bx^4+2cx^6
\end{equation}
for the Lagrangian in Eq.(27 )of the cubic-quintic Duffing oscillator. It is easy to combine Eqs. (65) and (66) to obtain  differential equation (25) thus verify that Eq. (66) indeed represent the correct Hamiltonian. A more interesting example in respect of this is provided by the Li\'{e}nard-type oscillator in Eq. (30) for which we wrote two Lagrangian functions, namely, the logarithmic and inverse-type expressions. These nonstandard Lagrangians are not connected by gauge terms. These two results, therefore, provide alternative analytic representation of the dynamical system. The Logarithmic type result in Eq. (33) when used in the definition of canonical momentum leads to an expression that does not permit us to express $\dot{x}$ as a function of $p$ and $x$, and is therefore unsuitable to give a Hamiltonian representation of the oscillator. Fortunately, this is not the case with the inverse-type Lagrangian of Eq. (34). Here use of Eq. (63) gives
\begin{equation}
 \dot{x}=\frac{\sqrt{3k}-(9\lambda+k^2x^2)\sqrt{p}}{3k\sqrt{p}}
\end{equation}
which in conjunction with (64) leads to the Hamiltonian
\begin{equation}
 H=\frac{2\sqrt{p}}{\sqrt{3k}}-\frac{p}{k}(\frac{1}{3}k^2x^2+3\lambda).
\end{equation}
Results similar to those in Eqs. (67) and (68) for the Lagangian in Eq. (37) of the Mathews-Lakshmanan  oscillator read 
\begin{equation}
 \dot{x}=\frac{(1+\lambda x^2)p}{2\lambda}
\end{equation}
and
\begin{equation}
 H=\frac{p^2}{4\lambda}(1+\lambda x^2)-\frac{\omega^2}{(1+\lambda x^2)^2}.
\end{equation}
The other Lagrangian in Eq. (38) is not suitable to provide a similar Hamiltonian representation of the system.    
\par  We have found indirect Lagrangian representations for the damped Harmonic oscillator, Lorentz- and Van der Pol oscillators. These systems can also be Hamiltonized by using the procedure followed above. For example, the well known results for canonical momenta and Hamiltonian of the damped Harmonic oscillator are given by \cite{8}
\begin{equation}
 p_x=\dot{y}-\frac{1}{2}\gamma y,\;\;\;\;p_y=\dot{x}+\frac{1}{2}\gamma x
\end{equation}
and
\begin{equation}
 H=p_xp_y+\frac{1}{2}\gamma(yp_y-xp_x)+(1-\frac{1}{4}\gamma^2)xy.
\end{equation}
Here $p_x$ and $p_y$ stand for canonical conjugate to $x$ and $y$ coordinates. The results for the canonical momenta and Hamiltonian for the Lorentz oscillator closely resemble those of the damped Harmonic oscillator and are given by 
\begin{equation}
 p_x=\dot{y}-\frac{1}{2}\delta y,\;\;\;\;p_y=\dot{x}+\frac{1}{2}\delta x
\end{equation}
and
\begin{equation}
 H=p_xp_y+\frac{1}{2}\delta(yp_y-xp_x)+\alpha x^2 y-(\mu+\frac{1}{4}\delta^2)xy.
\end{equation}
Similar results for the Van der Pol oscillator read
\begin{equation}
 p_x=\dot{y}+\frac{1}{2}\mu y-\frac{1}{4}\mu x^2y,\;\;\;\;p_y=\dot{x}-\frac{1}{2}\mu x+\frac{1}{4}\mu x^3
\end{equation}
and
\begin{equation}
 H=p_xp_y+\frac{1}{2}\mu(xp_x-yp_y)+\frac{1}{4}\mu x^2(yp_y-xp_x)+\frac{1}{4}\mu^2x^3y(1-\frac{1}{4}x^2)+(1-4\mu^2)xy.
\end{equation}
The Abraham-Lorentz equation (39) is characterized by a second-order Lagrangian. Consequently, it was Hamiltonized by Englert \cite{34} by applying Ostrogradsky formalism \cite{45} to generalized momenta. More recently, Bender et al \cite{46}, found a simpler quadratic Hamiltonian for the system.
\section*{5. Concluding remarks}
In the calculus of variation the so-called direct problem represents the conventional method in which one first assigns a Lagrangian to a physical system and then computes the equation of motion through Euler-Lagrange equations. Contrarily, the inverse problem consists in constructing the Lagrangian functions from the equation of motion. The inverse problem for systems described by linear differential equation has an old root in the classical mechanics literature \cite{47}. This is, however, not the case with systems modeled by nonlinear differential equations. Only in the recent past, there were attempts \cite{11,13,14} to preserve Lagrangian structure in the variational formulation of nonlinear equations. The nonlinear differential equations were found to admit, what we now call, nonstandard Lagrangian representation. These Lagrangians involve neither the kinetic energy nor the potential energy and were proposed to identify the class of equations that admit a Lagrangian description. For example, in ref. 11 a general form of Eq. (5a) written as $L=1/(\dot{x}+kU(x,t))$ was substituted in the Euler-Lagrange equation to verify that this proposed expression for $L$ stands for the Lagrangian for the second-order Riccati equation provided we choose $U(.)=c_0(t)+c_1(t)x+c_2(t)x^2$. In this work we have made use of two different methods to provide a complete solution for the inverse problem of the calculus of variations for systems modeled by nonlinear equations. In the first method we have chosen to work with an integral representation of the Lagrangian function in terms of the first integral $K(x,\dot{x})$ of the associated equation of motion. As a useful application of the method we provided only a few case studies. However, the method is quite general but depends crucially on the efficiency of the method used for finding the first integral of the equation of motion. One may try to improve on the method followed by us for computing results for $K(x,\dot{x})$. Such studies, on the one hand, are expected to broaden the scope of applicability of this approach to a wide variety of nonlinear differential equations and, on the other hand, are likely to shed new light on the systems' integrability \cite{48}. In the second method of our interest we adapted the approach derived in ref. 20 to deal with velocity dependent nonlinear equations although it was originally developed to treat the inverse problem of linear differential equations only. We computed results for the Jacobi integrals corresponding to Lagrangian functions found by both methods and used them to provide appropriate Hamiltonian representations of the systems without taking recourse to the use of so-called Legendre transformation \cite{7}.
\par Although our main objective in this work was to make use of some uncomplicated method to solve the inverse of the calculus of variation for nonlinear equations, we provided here a case study for a linear third order system (Abraham-Lorentz equation) with the hope that the result presented may motivate variational studies in higher-order nonlinear systems. In this case the solution of the inverse problem led to a second-order Lagrangian for which we made use of Eq. (56) to obtain the Jacobi integral in Eq. (62). We did not make any attempt to obtain the Hamiltonian from the Jacobi integral presumably because the Hamiltonian function for Abraham-Lorentz equation has been given elsewhere \cite{45,46}.      
\par The inverse problems related to natural sciences, ranging from geophysics to medical diagnostics \cite{49}, have been widely discussed in the literature. Formally, to find the solution of an inverse problem amounts to discovering the cause of observed data. In classical mechanics, the solution of the inverse problem essentially consists in expressing an ordinary differential equation (linear or nonlinear) in Hamilton's variational form. It remains an interesting curiosity to extend our treatment to field theory \cite{50} where dynamics of physical systems are expressed by partial differential equations.


\vskip 1cm
\begin{thebibliography}{99}
\bibitem{1} Santilli  R M 1978 {\it Foundation of Theoretical Mechanics}, Vol. 1- {\it The inverse Problem in Newtonian Mechanics}, Springer Verlag, New York
\bibitem{2} Douglas J 1941 Trans. Amer. Math. Soc. {\bf 50}, 71 
\bibitem{3} Caldirola  P 1941 Il Nuovo Cimento {\bf 18}, 393 
\bibitem{4} Kanai E 1948 Prog. Theor. Phys. {\bf 3}, 440
\bibitem{5} Baldiotti M C, Fresneda R and Gitman D M 2011 Phys. Lett. A {\bf 375}, 1630
\bibitem{6} Bateman  H 1931  Phys. Rev. {\bf 38}, 815
\bibitem{7} Goldstein H 1950 {\it Classical Mechanics}, Addison-Wesley, Reading, MA
\bibitem{8} Blasone M and Jizba P 2004  Ann. Phys. (NY) {\bf 312}, 354 
\bibitem{9} Takahashi K 2018 J. Math. Phys. {\bf 59}, 032103 
\bibitem{10} Takahashi K 2018 J. Math. Phys. {\bf 59}, 072108 
\bibitem{11} Cari\~{n}ena J F, Ra\~{n}ada M F and Santander F 2005 J. Math. Phys. {\bf 46}, 062703 
\bibitem{12} Chandrasekar V K, Senthivelan M and Lakshmanan M 2007 J. Math. Phys. {\bf 48}, 032701
\bibitem{13} Musielak Z E 2008 J. Phys. A: Math. Theor. {\bf 41}, 055205
\bibitem{14} Cie\'{s}li\'{n}ski J L and Nikiciuk T 2010 J. Phys. A: Math. Theor. {\bf 43}, 175205 
\bibitem{15} Ince E L 1958 {\it Ordinary Differential Equations}, Dover Publications, New York 
\bibitem{16} Saha A and Talukdar B 2014 Rep. Math. Phys. {\bf 73}, 299
\bibitem{17} Nucci M C and Tamizhmani K M 2010  J. Nonlin. Math. Phys. {\bf 17}, 167
\bibitem{18} Whittaker E T 1988 {\it A Treatise on the Analytical Dynamics of Particles and Rigid Bodies}, Cambridge University Press, Cambridge,  First published, 1904
\bibitem{19} L\'{o}pez G 1996 Ann. Phys. (NY) {\bf 251}, 363 
\bibitem{20} Hojman S A 1984 J. Phys. A: Math. Gen. {\bf 17}, 2399 
\bibitem{21} Olver P J 1993 {\it Application of Lie groups to differential equations}, 1st ed. Springer, New York
\bibitem{22} Ghosh Subrata, Shamanna J and Talukdar B 2004 Can. J. Phys. {\bf 82}, 561
\bibitem{23} Morandi G, Ferrario C, Vecchio G Lo, Marmo G and Rubano C 1990 Phys. Rep. {\bf 188}, 1
\bibitem{24} Hojman R, Hojman S A and Sheinbaum J 1983 Phys. Rev. D {\bf 28}, 1333 
\bibitem{25} Currie D G and Saletan E J 1966 J. Math. Phys. {\bf 7}, 967
\bibitem{26} Courant R and Hilbert D 1975 {\it Methods of Mathematical Physics}, Vol. 1 Wiley Eastern Pvt. Ltd., New Delhi
\bibitem{27} Talukdar B and Das U 2008 {\it Higher-Order Systems in Classical Mechanics}, Norasa Publishing House, New Delhi
\bibitem{28} Caratheodory C 1967 {\it Calculus of variations and partial differential equations of first order}, Vol. 2, Second ed. Holden Day, San Fransisco
\bibitem{29} El\'{\i}as-Z\'{u}\'{n}iga A 2013  App. Math. Modelling {\bf 37}, 2574 
\bibitem{30} Mathews P M and Lakshmanan M 1975 Il Nuovo Cimento A {\bf 26}, 299 
\bibitem{31} Jackson J D 2011 (reprint) {\it Classical Electrodynamics}, Nice Printing press, Delhi, India 
\bibitem{32} Almendral J A and Sanjuan M A F 2003  J. Phys. A: Math. Gen. {\bf 36}, 695 
\bibitem{33} Van der Pol B and Van der Mark J 1927 Nature {\bf 120}, 363 
\bibitem{34} Englert B 1980 Ann. Phys. (NY) {\bf 129}, 1 
\bibitem{35} Nayfeh A H and Mook D T 1995 {\it Nonlinear oscillations}, John Wiley, New York
\bibitem{36} Chandrasekar V K, Senthilvelan M and Lakshmanan M 2005 Phys. Rev. E {\bf 72}, 066203 
\bibitem{37} Delbourgo R, Salam A and Strathdee 1969 Phys. Rev. {\bf 187}, 1999 
\bibitem{38} Koc R and and Koca M 2003 J. Phys. A: Math. Gen. {\bf 36}, 8105   
\bibitem{39} Lakshmanan M and Chandrasekar V K 2013 Eur. Phys. J. Spec. Top. {\bf 222}, 665 
\bibitem{40} Feng Z, Gao G and Cui J 2011 Comm. Pure and Appl. Analysis {\bf 10}, 1377
\bibitem{41} Ghosh S, Talukdar B and Sarkar P 2007 Acta Mechanica {\bf 190}, 73 
\bibitem{42} Schiff L I 2010 {\it Quantum Mechanics}, Tata McGraw-Hill Edition, New Delhi 
\bibitem{43} Talukdar Benoy, Chatterjee Supriya and Golam Ali Sekh, On the analytic representation of Newtonian systems, arXiv: 2006.08597v1 [physics.class-ph] 12 Jun 2020
\bibitem{44} Shamanna J, Talukdar B and Das U 2002 Phys. Lett. A {\bf 305}, 93 
\bibitem{45} Ostrogradsky M 1850 Mem. Ac. St. Petersbourg {\bf VI}, 385 
\bibitem{46} Bender C M, Gianfreda M, Hassanpour N and Jones H F 2016 J. Math. Phys. {\bf 57}, 084101
\bibitem{47} Helmholtz H 1887 Journal f\"{u}r die reine und angewandte Mathematik {\bf 100}, 137 
\bibitem{48} Stachowiak T, Hypergeometric First Integrals of the Duffing and van der Pol Oscillators, arXiv: 1706.02506v3 [Math-ph] 15 Nov 2018
\bibitem{49} Tikhonov A N and Goncharsky A V (Ed.) 1987 {\it Ill-Posed Problems in the Natural Sciences} (Mathematics and Mechanics Series), Translated by M. Bloch, MIR Publisher, Moscow
\bibitem{50} Malvern L E 1969, {\it Introduction to the mechanics of a continuous medium}, Prentice-Hall, Inc., New Jersey 
\end{thebibliography}
\end{document}